\documentclass{conm-p-l}

\copyrightinfo{2017}{}

\usepackage{graphicx}
\usepackage{subfigure}

\usepackage{amssymb,amsmath,amsthm,amscd}

\theoremstyle{definition}

\theoremstyle{remark}

\numberwithin{equation}{section}

\newcommand{\Dl}{\Delta}

\newcommand{\sg}{\sigma}

\newcommand{\pa}{\partial}

\newcommand{\na}{\nabla}

\begin{document}

\title[Novel discoveries]{Novel discoveries on the mathematical foundation of linear hydrodynamic stability theory}

\author{Y. Charles Li}
\address{Y. Charles Li, Department of Mathematics, University of Missouri, 
Columbia, MO 65211, USA}
\email{liyan@missouri.edu}
\urladdr{http://faculty.missouri.edu/~liyan}

\curraddr{}
\thanks{}

%\subjclass{Primary 76, 35; Secondary 34}

\subjclass{PACS: 47.10.-g; 47.27.-i}

\date{}

\dedicatory{}

\keywords{Hydrodynamic stability, Short term unpredictability, rough dependence on initial data, turbulence, super fast growth}

\begin{abstract}
We present some new discoveries on the mathematical foundation of linear hydrodynamic stability theory. The new discoveries are: 1. Linearized Euler equations fail to provide a linear approximation on inviscid hydrodynamic stability. 2. Eigenvalue instability predicted by high Reynolds number linearized Navier-Stokes equations cannot capture the dominant instability of super fast growth. 3. As equations for directional differentials, Rayleigh equation and Orr-Sommerfeld equation cannot capture the nature of the full differentials. 
\end{abstract}

\maketitle

\section{Introduction}

Hydrodynamic stability goes back to such luminaries as Helmholtz, Kelvin, Rayleigh, O. Reynolds, A. Sommerfeld, Heisenberg, and G. I. Taylor. Many treatises have been written on the subject \cite{Lin55} \cite{Cha61} \cite{Dra02}.

Linear hydrodynamic stability theory is based upon the simple idea of linear approximation of a function through its tangent line. On the other hand, linear hydrodynamic stability theory deals with function spaces rather than one-dimensional Euclidean space. So justifying the simple idea for linear hydrodynamic stability theory is not simple. Hydrodynamicists have been trying to lay down a rigorous mathematical foundation for linear hydrodynamic stability theory ever since its beginning \cite{Lin55}. But a rigorous mathematical foundation can only be achieved with enough advances of rigorous PDE results on Navier-Stokes (or Euler) equations. Recent advances on the solution operators of Euler and Navier-Stokes equations \cite{Inc15} \cite{Li14} \cite{Li17} \cite{FL17} provided enough base to evaluate the rigorous mathematical foundation of linear hydrodynamic stability theory. We discovered that linear Euler equations fail to provide a linear approximation to inviscid hydrodynamic stability. The simple tangent line approximation does not apply here due to the fact that the derivative does not exist. Even though linearized Navier-Stokes equations can provide a linear approximation to viscous hydrodynamic stability, initial stage dynamics of high Reynolds number linear Navier-Stokes equations are often dominated by amplifications much faster than exponential growths. Such super fast amplifications of perturbations can quickly reach substantial magnitudes so that nonlinear effects are substantial (out of the linear regime). Thus exponential growths generated by unstable eigenvalues to high Reynolds number linear Navier-Stokes equations often cannot be observed at all. For the channel flow, the Rayleigh equation and Orr-Sommerfeld equation can be viewed as equations of directional derivatives of the solution  operators to Euler and Navier-Stokes equations. It is well-known that directional derivatives cannot capture the nature of the full derivatives (gradients). In the current case, Rayleigh equation predicts an exponential stability/instability, while the full derivative of the solution operator to Euler equations  does not exist \cite{Inc15} and the temporal amplifications of perturbations can be much faster than exponential growths. The Orr-Sommerfeld equation predicts an exponential stability/instability, while the full derivative of the solution operator to Navier-Stokes equations can lead to temporal amplifications of perturbations, which are much faster than exponential growths.

\section{Linear approximation in the simplest setting}

Let us start with a real valued function of one variable
\[
y = f(x).
\]
If the function is differentiable, then
\[
\Dl y = f'(x) \Dl x  + o(|\Dl x|).
\]
Thus in a small neighborhood of $\Dl x = 0$, the tangent line
\[
dy = f'(x) dx
\]
offers a linear approximation. In fact, sometimes there is a nice change of variables
\[
\Dl \eta = \Dl y + g(\Dl y),
\]
where $g(0)=g'(0)=0$ such that
\[
\Dl \eta = f'(x) \Dl x
\]
in a small neighborhood of $\Dl x = 0$. This is called linearization. The same idea can be applied to differential equations as well. Let us consider a simple first order equation,
\[
\frac{dx}{dt} = f(x).
\]
Denote by $S^t(x_0)$ the solution operator
\[
x(t) = S^t(x_0)
\]
where $x(0) = x_0$. The solution operator can be viewed as a family of maps parametrized by $t$. If $S^t(x_0)$ is differentiable in $x_0$, then
\[
\Dl x (t) = \frac{d}{dx_0} S^t(x_0) \Dl x_0 +o (|\Dl x_0|).
\]
In a small neighborhood of $\Dl x_0 = 0$, the tangent line
\[
dx(t) = \frac{d}{dx_0} S^t(x_0) dx_0
\]
offers a linear approximation. $dx(t)$ satisfies the linear equation
\[
\frac{d}{dt} (dx)= f'(x) dx,
\]
whereas $\Dl x$ satisfies
\[
\frac{d}{dt} (\Dl x)= f'(x) \Dl x +o(|\Dl x|).
\]
Sometimes, there is a nice change of variables
\[
\Dl \eta = \Dl x + g(\Dl x)
\]
where $g(0)=g'(0)=0$ such that
\[
\frac{d}{dt} (\Dl \eta )= f'(x) \Dl \eta
\]
in a small neighborhood of $\Dl \eta =0$. This is called linearization of the differential equation.

\section{Mathematical foundation of inviscid linear hydrodynamic stability theory}

We start with the incompressible Euler equations
\[
\pa_t u +u \cdot \na u = - \na p , \  \na \cdot u = 0
\]
under various boundary conditions, where the spatial dimension is either $2$ or $3$. In order to establish linear hydrodynamic stability theory, first of all the Euler equations must be well-posed. The usual well-posedness requires that there is a unique solution which depends on the initial condition continuously. In fact, linear hydrodynamic stability theory requires a stronger well-posedness, i.e. the solution needs to depend on the initial condition differentiably as discussed later. The well-posedness of Euler equations depends critically on the function space in which the Euler equations are posed. First of all, the function space consists of functions that satisfy the boundary condition associated with the Euler equations. Second, a norm is defined in the function space so that every function with a finite norm belongs to the function space. The most natural function spaces for the well-posedness of Euler equations are the Sobolev spaces $H^n$ ($n$ is a positive integer). Besides satisfying the boundary condition associated with the Euler equations, a function belonging to $H^n$ must have a finite Sobolev norm which is the square root of the integral over the spatial domain of the sum of each velocity component's square and each velocity component's partial derivative (up to order $n$) square, i.e. sum of
\[
\left [ \left ( \frac{\pa}{\pa x_1} \right )^\ell   \left ( \frac{\pa}{\pa x_2} \right )^m u_j \right ]^2, \
(0 \leq \ell , m \leq n, \ \ell + m  \leq n, \ j=1,2)
\]
in two dimensions, similarly for three dimensions. The boundary conditions define different types of fluid flows. Under the boundary conditions of either decaying at infinity of the whole space or spatial periodicity in all the axial directions, mathematical analysis proves that the Euler equations are locally well-posed in the Sobolev spaces $H^n$ when $n > 1 + \frac{d}{2}$ ($d$ is the spatial dimension) \cite{Kat72} \cite{Kat75}. Here local well-posedness means that during a short time depending on the norm of the initial condition, it is well-posed. In two spatial dimensions, the local well-posedness can be extended to global (i.e. for all time) well-posedness. There are also well-posedness results for other boundary conditions, and boundary conditions usually do not pose any substantial issue for well-posedness. Once we know well-posedness, we can define a flow in the function space. For any initial condition in $H^n$ ($n > 1+\frac{d}{2}$), the corresponding solution draws an orbit in $H^n$ either locally or globally in time. The visual picture is that an initial condition as any function in $H^n$ can be viewed as a
"point" in $H^n$, and the solution starting from the initial condition can be viewed as an "orbit" in $H^n$. In fact, we can define a solution operator $S^t$ on $H^n$ for Euler equations as follows:
\[
S^t(u(0)) = u(t), \   \    \text{for any } u(0) \in H^n,
\]
where $u(t)$ is the solution starting from the initial condition $u(0)$. The well-posedness of Euler equations in $H^n$ implies that $S^t(u(0))$ is continuous in $u(0)$. To study the stability of any solution $u(t)$, one needs to introduce a perturbation to its initial condition
\[
u(0) + \Dl u(0)
\]
which generates a new solution to Euler equations
\[
\hat{u} (t) = S^t(u(0)+\Dl u(0)).
\]
The stability of the solution $u(t)$ is studied via investigating the growing or decaying nature of
\begin{eqnarray*}
\Dl u(t) &=& \hat{u} (t) - u(t) \\
&=& S^t(u(0)+\Dl u(0)) - S^t(u(0)).
\end{eqnarray*}
The linear hydrodynamic stability theory must be based on the linear approximation to $\Dl u(t)$ as a function of $\Dl u(0)$, which requires that $S^t(u(0))$ is differentiable in $u(0)$. If $S^t(u(0))$ were differentiable in $u(0)$, then we could have
\begin{equation}
\Dl u(t) = \left [ \na_{u(0)} S^t(u(0)) \right ] \Dl u(0) + o \left ( \| \Dl u(0) \|_n \right ) , \label{DA}
\end{equation}
where $\na_{u(0)}$ represents the derivative in $u(0)$, and $\| \   \|_n$ is the $H^n$ norm. In a small neighborhood of $\Dl u(0)=0$, we could have the linear approximation
\begin{equation}
d u(t) = \left [ \na_{u(0)} S^t(u(0)) \right ] d u(0) .\label{LA1}
\end{equation}
In fact, $du(t)$ would satisfy the linearized Euler equations
\begin{equation}
\pa_t d u(t) + du \cdot \na u + u \cdot \na du= -\na dp, \   \na \cdot du = 0 .\label{LA2}
\end{equation}
Unfortunately, recent mathematical analysis proved that $S^t(u(0))$ is nowhere differentiable in either $2$ or $3$ spatial dimensions under decaying boundary condition at infinity \cite{Inc15}. That is, at any initial condition in $H^n$, the solution operator $S^t(u(0))$ is not differentiable in $u(0)$.  It is believed that the claim should hold for other boundary conditions as well \cite{Li14}. The claim should be a generic fact associated with Euler equations. The derivative (\ref{DA}) never exists, the differential (\ref{LA1}) never exists, and the linearized Euler equations (\ref{LA2}) always fail to provide a linear approximation for the inviscid hydrodynamic stability. In conclusion, the inviscid linear hydrodynamic stability theory always fails to have a rigorous mathematical foundation!

There are different ways for the non-existence of the derivative (\ref{DA}). The most common way is that the norm of the derivative $\na_{u(0)} S^t(u(0))$ is infinite as shown by the explicit example in \cite{Li17}. In such a case, the amplification from $\Dl u(0)$ to $\Dl u(t)$ can be expected to be super fast, and much faster than the exponential growth associated with any unstable eigenvalue of the linearized Euler equations (\ref{LA2}) \cite{Li14} \cite{Li17}. This is the phenomenon that we call "rough dependence upon initial data" (also "super fast amplification" or "short term unpredictability") \cite{Li14} \cite{Li17}. Such super fast growth can reach substantial magnitude in very short time during which the exponential growth is very small. Thus the classical unstable eigenvalue inviscid instability does not capture the true inviscid instability of super fast growth.

\section{Rayleigh equation as an equation for a directional differential}
Now we focus on the channel flow, for example, a 2D channel flow: $x_1$ is the streamwise direction and $x_2$ is the transverse direction with the boundaries at $x_2 =a, b$. The classical interest is to study the linear stability of the steady state
\[
u_1 = U(x_2), \   u_2 = 0 ,
\]
which always satisfies the boundary condition
\[
u_2 = 0 , \   \   \text{at } x_2 = a, b.
\]
As discussed before, it is believed that the solution operator is not differentiable at the steady state, so the linearized Euler equations at the steady state cannot guarantee a linear approximation. Nevertheless, the classical inviscid linear hydrodynamic stability theory was based upon the false assumption that the solution operator is differentiable. Formally starting from the linearized Euler equations, Rayleigh derived the so-called Rayleigh equation as follows. Let
\begin{equation}
du = (\pa_{x_2} \psi , - \pa_{x_1} \psi ), \  \psi = \phi (x_2) e^{ik(x_1 - \sg t)} , \label{DD1}
\end{equation}
then $\phi (x_2)$ satisfies the Rayleigh equation
\begin{equation}
(U-\sg ) (\phi ''  - k^2 \phi ) - U'' \phi = 0 , \label{RE}
\end{equation}
with the boundary condition
\[
\phi (a) = \phi (b) = 0.
\]
Setting $t=0$ in (\ref{DD1}), we have
\begin{equation}
du (0)= (\pa_{x_2} \psi (0), - \pa_{x_1} \psi  (0)), \  \psi (0)= \phi (x_2) e^{ikx_1} . \label{DD2}
\end{equation}
One can view the $du(0)$  in (\ref{DD2}) as a single Fourier mode in $x_1$ out of the whole Fourier series in $x_1$. In fact, (\ref{DD1}) represents a directional differential with the direction of $du(0)$ specified by $k$ and $\phi (x_2)$. Even though the full differential $du(t)$ does not exist, the directional differential (\ref{DD1}) can still exist once the Rayleigh equation (\ref{RE}) produces an eigenfunction. Thus the directional differential (\ref{DD1}) generated from the Rayleigh equation (\ref{RE}) cannot capture the nature of the full differential (\ref{LA1}) which does not exist. As mentioned before, the most common way for the non-existence of (\ref{LA1}) is that the norm of the derivative $\na_{u(0)} S^t(u(0))$ is infinite. This will result in super fast growth of certain perturbations $\Dl u(0)$, which is much faster than any exponential growth predicted by Rayleigh equation. Such super fast growth can reach substantial magnitude in very short time. The super fast growing perturbations are not the directional perturbation (\ref{DD1}) of Rayleigh. Instability is usually dominated by the fastest growing perturbation. Thus Rayleigh equation cannot capture the dominant inviscid instability of super fast growth which can reach substantial magnitude in very short time - short term unpredictability.

\section{High Reynolds number linear hydrodynamic stability theory}

We start with the incompressible Navier-Stokes equations
\[
\pa_t u + u \cdot \na u = - \na  p +\frac{1}{Re} \na^2 u , \   \na \cdot u = 0
\]
under various boundary conditions, where the spatial dimension is either $2$ or $3$. Under the boundary conditions of either decaying at infinity of the whole space or spatial periodicity in all the axial directions, like Euler equations, Navier-Stokes equations are locally well-posed in the Sobolev space $H^n$ when $n > 1 + \frac{d}{2}$ ($d$ is the spatial dimension) \cite{Kat72} \cite{Kat75}. In two spatial dimensions, the local well-posedness can be extended to global well-posedness. Such well-posedness can also be established under other boundary conditions. Like for Euler equations, the solution operator $S^t(u(0))$ for Navier-Stokes equations is defined as follows:
\[
S^t(u(0)) = u(t), \   \    \text{for any } u(0) \in H^n,
\]
where $u(t)$ is the solution to Navier-Stokes equations, with the initial condition $u(0)$. Unlike for Euler equations, the solution operator $S^t(u(0))$ for Navier-Stokes equations is not only  continuous but also everywhere differentiable in $u(0)$. Then we have
\begin{equation}
\Dl u(t) = \left [ \na_{u(0)} S^t(u(0)) \right ] \Dl u(0) + o \left ( \| \Dl u(0) \|_n \right ) .
\end{equation}
In a small neighborhood of $\Dl u(0)=0$, we have the linear approximation
\begin{equation}
d u(t) = \left [ \na_{u(0)} S^t(u(0)) \right ] d u(0) .\label{NSLA}
\end{equation}
As a result, $du(t)$ satisfies the linearized Navier-Stokes equations
\begin{equation}
\pa_t d u(t) + du \cdot \na u + u \cdot \na du= -\na dp + \frac{1}{Re} \na^2 du, \   \na \cdot du = 0 .\label{NSLE}
\end{equation}
The norm of the derivative $\na_{u(0)} S^t(u(0))$ as a map which maps $du(0)$ to $du(t)$, is defined as
\[
\left \| \na_{u(0)} S^t(u(0)) \right \| = \sup_{du(0)} \frac{\| du(t) \|_n} {\| du(0) \|_n}  .
\]
Under either decaying at infinity or periodic boundary condition in both two and three spatial  dimensions, the norm of the derivative $\na_{u(0)} S^t(u(0))$ is bounded as follows \cite{Li14}
\begin{equation}
\left \| \na_{u(0)} S^t(u(0)) \right \| \leq e^{\sg \sqrt{Re} \sqrt{t} + \sg_1 t}, \  t \in [0, T] \label{DBD}
\end{equation}
where
\[
\sg_1 = \frac{\sqrt{2e}}{2} \sg , \  \sg = \frac{8}{\sqrt{2e}} \sup_{\tau \in [0, T]} \| u(\tau )\|_n ,
\]
and [$0, T$] is a time interval on which well-posedness holds. The exponent of the bound has two parts, the first part depends on the square root of $Re$ and $t$, and the second part is independent of $Re$. As $Re$ approaches infinity, the bound also approaches infinity in agreement with the intuition that the norm of the derivative approaches its inviscid counterpart. The peculiar feature of the first part is the square root nature. At $t=0$, the time derivative of the first part is infinite. During the time interval $t \in (0, \frac{2}{e} Re )$, the first part is greater than the second part. When the Reynolds number $Re$ is large, the time interval is very large. When the Reynolds number $Re$ is large, the first part is very large even during short time, for example $t \in (0, \frac{1}{\sqrt{Re}})$. Thus, when the Reynolds number $Re$ is large, the first part corresponds to super fast growth. We also call such super fast growth, "rough dependence upon initial data" (in viscous case) or "short term unpredictability" (in viscous case). The second part corresponds to the unstable eigenvalue exponential growth. Examples show that as the Reynolds number approaches infinity, the viscous unstable eigenvalues approach the corresponding inviscid unstable eigenvalues \cite{Li05}.That is why the second part does not depend on the Reynolds number.  When the Reynolds number $Re$ is large, the super fast growth due to the first part can reach substantial magnitude in very short time, e.g. $t \in (0, \frac{1}{\sqrt{Re}})$ during which the exponential growth due to the second part is very small
$e^{\frac{\sg_1}{\sqrt{Re}}}$. Thus the eigenvalue instability cannot capture the dominant instability of super fast growth, when the Reynolds number is large.

Numerical simulations on the linearized Navier-Stokes equations (\ref{NSLE}) under periodic boundary condition are conducted for different base solutions and different perturbations \cite{FL17}. The $e^{c \sqrt{t}}$ nature in (\ref{DBD}) is indeed observed for the amplification of abundant perturbations (\ref{NSLA}) along abundant base solutions \cite{FL17}. The choice of the base solution and perturbation is decided by the initial conditions $u(0)$ and $du(0)$. Our conclusion is that high Reynolds number fully developed turbulence is manifested via constant super fast amplifications of ever existing perturbations.

\section{Orr-Sommerfeld equation as an equation for a directional differential}

In the case of channel flow, the viscous counterpart of the Rayleigh equation is the Orr-Sommerfeld equation which can be derived by substituting (\ref{DD1}) into the linearized Navier-Stokes equations (\ref{NSLE}). The Orr-Sommerfeld equation takes the form
\begin{equation}
\phi '''' - 2k^2 \phi '' + k^4 \phi = ik Re [(U-\sg ) (\phi ''  - k^2 \phi ) - U'' \phi ] , \label{OSE}
\end{equation}
with the boundary condition
\[
\phi (a) = \phi (b) = \phi '(a) = \phi ' (b) = 0.
\]
Even though in the viscous case, the differential (\ref{NSLA}) exists and satisfies the linearized Navier-Stokes equations (\ref{NSLE}), the directional differential (\ref{DD1}) generated by the eigenfunction of the Orr-Sommerfeld equation (\ref{OSE}) cannot capture the nature of the full differential (\ref{NSLA}). The instability predicted by the unstable eigenvalues of the Orr-Sommerfeld equation, is the exponential growth described by the second part  in the exponent of the bound (\ref{DBD}). On the other hand, the norm of the full derivative can have a super fast growing factor described by the first part in the exponent of (\ref{DBD}), and there are abundant initial perturbations $du(0)$ that lead to such super fast amplifications as shown numerically \cite{FL17}. The super fast amplifying perturbations are not the directional perturbation (\ref{DD1}) generated by the eigenfunction of the Orr-Sommerfeld equation. Instability is usually dominated by the fastest growing perturbations. Thus, when the Reynolds number is large, exponential instability predicted by the unstable eigenvalues of the Orr-Sommerfeld equation cannot capture the dominant instability of the super fast growth which can reach substantial magnitude in very short time - short term unpredictability.

\section{Conclusion}

We discover that linear Euler equations fail to provide a linear approximation to inviscid hydrodynamic stability. Linear inviscid hydrodynamic instability is dominated by super fast growths of perturbations which can reach substantial magnitude in very short time during which the exponential growth due to unstable eigenvalues of linear Euler equations is very small. As an equation for a directional differential, Rayleigh equation can predict unstable eigenvalue exponential instability which cannot capture the dominant inviscid linear instability of super fast growth. Even though linearized Navier-Stokes equations can provide a linear approximation to viscous hydrodynamic stability, the exponential growth due to unstable eigenvalues of the linearized Navier-Stokes equations cannot capture the dominant linear instability of super fast growth when the Reynolds number is large. The super fast growth can reach substantial magnitude in a very short time during which the exponential growth is very small. As an equation for a directional differential, Orr-Sommerfeld equation can predict exponential instability due to unstable eigenvalues, and such exponential growth cannot capture the super fast growth of the full linearized Navier-Stokes equations when the Reynolds number is large.

Hydrodynamicists have been pursuing for a rigorous mathematical foundation of the linear hydrodynamic stability theory ever since its beginning \cite{Lin55}. Only the recent mathematical results provide enough base to clearly evaluate the mathematical foundation.

\end{document}